\documentclass[prc,twocolumn,showpacs]{revtex4}
\usepackage{amsmath}
\usepackage{epsfig}

\newcommand{\nslash}{\kern 0.2 em n\kern -0.50em /}

\newcommand{\beq}{\begin{eqnarray}}
\newcommand{\eeq}{\end{eqnarray}}

\def\bq{\begin{eqnarray}}
\def\eq{\end{eqnarray}}

\def\roughly#1{\mathrel{\raise.3ex\hbox{$#1$\kern-.75em
\lower1ex\hbox{$\sim$}}}}
\def\bm#1{\mbox{\boldmath$#1$\/}}

\begin{document}

\preprint{\hfill\parbox[b]{0.3\hsize}
{ }}

\def\bra{\langle }
\def\ket{\rangle }

\title{
Extracting generalized neutron parton distributions from $^3$He data 
}
\author{M. Rinaldi\footnote{Electronic address:
matteo.rinaldi@pg.infn.it},
S. Scopetta\footnote{Electronic
address: sergio.scopetta@pg.infn.it}}
\affiliation
{\it
Dipartimento di Fisica, Universit\`a degli Studi di Perugia, 
and INFN sezione di Perugia, via A. Pascoli 06100 Perugia, Italy
}

\begin{abstract}
 
An impulse approximation (IA) analysis is described of 
the generalized parton distributions (GPDs) $H$ and $E$ 
of the $^3$He nucleus,
quantities which are accessible
in hard exclusive processes, such as 
coherent deeply virtual
Compton scattering (DVCS).
The calculation is based on the Av18 interaction.
The electromagnetic form factors are correctly recovered
in the proper limits.
The sum of the GPDs $H$ and $E$ of $^3$He,
at low momentum transfer, is largely
dominated by the neutron contribution, thanks to
the unique spin structure of $^3$He.
This nucleus is therefore very promising
for the extraction of the neutron information. 
By increasing the momentum transfer, however,
this conclusion is somehow hindered by the  
the fast growing proton contribution.
Besides, even when 
the neutron contribution to the GPDs of $^3$He is largely dominating,
the procedure of extracting the neutron GPDs
from it could be, in principle, nontrivial.
A technique is therefore proposed, independent on both the
nuclear potential and the nucleon model used in the calculation,
able to take into account the nuclear effects included in the IA analysis
and to safely extract the neutron information at values of the
momentum transfer large enough to allow the measurements. 
Thanks to this observation, coherent DVCS
should be considered a key experiment 
to access 
the neutron GPDs and, in turn,
the orbital angular momentum of the partons in the neutron.

\end{abstract}

\maketitle

\section{Introduction}

Generalized Parton Distributions (GPDs) 
\cite{Mueller:1998fv,Radyushkin:1996nd,Ji:1996ek}
parametrize the non-perturbative hadron structure
in hard exclusive processes,
encoding therefore a wealth of information.
For example, the hadron three-dimensional
structure \cite{Burkardt:2000za} and the parton
total angular momentum could be unveiled by the measurement
of GPDs, which
will be therefore a major achievement for Hadronic Physics
in the next few years.
In particular, in order to access the hadron angular momentum content,
from which the orbital angular momentum (OAM) part could be
estimated by subtracting the helicity one, measurable in
deep inelastic scattering inclusive (DIS) and semi-inclusive (SiDIS)
processes, the knowledge of two GPDs is mandatory \cite{Ji:1996ek}.
They are the two chiral even, parton helicity independent GPDs 
occurring at leading twist, i.e., 
$H$, a target helicity-conserving quantity,
and $E$, a target helicity-flip one.
The cleanest experiment to access them
is Deeply Virtual Compton Scattering (DVCS),
i.e. the process
$
e H \longrightarrow e' H' \gamma
$ when
$Q^2 \gg M^2$
(everywhere in this paper,
$Q^2=-q \cdot q$ is the momentum transfer between the leptons $e$ and $e'$,
$\Delta^2$ the one between the hadrons $H$ and $H'$
and $M$ is the nucleon mass)
\cite{Ji:1996ek,Vanderhaeghen:1999xj}.
Despite severe difficulties related to the complicated way
GPDs enter the measured cross sections, 
DVCS data for proton and nuclear targets 
are being analyzed
(recent results can be found in
Refs. \cite{hermes}) and
GPDs are being extracted
(see Refs. \cite{Guidal:2010de} and references therein). 

The relevance of measuring GPDs for nuclear targets
has been addressed in several papers, {starting 
from Ref. \cite{cano1}, where 
the formalism for the deuteron target was presented.
Soon after, DVCS off 
spin 0, $1/2$ and 1  nuclear targets
has been detailed in Ref. \cite{kir_mue}, and the impulse approximation (IA)
approach to DVCS off spin 0 nuclei has been discussed in Ref. \cite{guzey1}.
Microscopic calculations have been described
for the GPDs of the deuteron in Ref. \cite{cano2}, and for
spin zero nuclei in Ref. \cite{liuti1};
a possible link of nuclear GPDs measurements to
color transparency phenomena has been presented in Ref.
\cite{liuti2}.
One of the main motivations for addressing nuclear GPDs measurements
is the possibility of
distinguishing medium modifications of the structure
of bound nucleons from conventional Fermi motion
and binding effects, a cumbersome separation
in data collected through standard DIS experiments.
Off-shell effects have been studied in Ref. \cite{liuti3}
and different medium modifications of nucleon GPDs 
have been illustrated in Ref. \cite{modif}. The possibilities
offered by heavy nuclear targets and the effects
in nuclear matter are also being investigated \cite{nm}.

The experimental study of nuclear GPDs
could therefore seriously contribute to shed some 
light on the origin of the so called EMC effect \cite{emc},
a puzzle still far to be solved.}
Great attention has anyway to be paid to avoid confusing
unusual effects with
conventional ones.
To this respect, 
few-body nuclear targets,
for which realistic studies 
are possible and exotic effects
are in principle distinguishable,
play a special role.
To this aim,
in Ref. \cite{io}, an IA calculation
of $H_q^3$, the GPD 
of $^3$He
corresponding to the flavor $q$,
has been presented,
valid for $-\Delta^2 \ll Q^2,M^2$, {and in particular,
for $|\Delta^2| \lesssim 0.3$ GeV$^2$}.
The approach permits to investigate 
the coherent, no break-up channel of DVCS off $^3$He,
{whose cross-section,  
at $|\Delta^2| \simeq 0.3$ GeV$^2$, is already too small
to be measured at present facilities}.
The main conclusion
was that the nuclear GPDs cannot be trivially
inferred from those of nuclear parton distributions (PDFs),
measured in DIS.

In a recent Rapid Communication of ours, Ref. \cite{nostro}, 
the approach of Ref. \cite{io} has been extended 
to evaluate the GPD $E_q$ of $^3$He, $E_q^3$.
The main goal was to study the possibility of accessing
the neutron information, which is very relevant because
it permits, together with the proton one, a flavor decomposition of 
GPDs data. One should not forget
that the properties of the free neutron 
have to be investigated using nuclear targets, 
taking nuclear effects properly into account.
In particular $^3$He, thanks to its peculiar spin structure,
has the unique property of simulating an effective polarized free
neutron target (see, e.g., \cite{friar,antico,SS}).
$^3$He is therefore a serious candidate
to measure the polarization properties
of the free neutron, such as its helicity-flip GPD $E_q$.
In Ref. \cite{nostro} it has been found that the sum of the GPDs
$H_q^3$ and $E_q^3$,
at low momentum transfer, is indeed 
dominated to a large extent by the neutron contribution,
making $^3$He targets very promising
for the extraction of the neutron information.
However, this is not the end of the story.
The same analysis has shown in fact that 
the proton contribution grows fast with increasing the momentum transfer.
Besides, even if 
the neutron contribution to the GPDs of $^3$He were largely dominating,
the procedure of extracting the neutron GPDs
from it could be, in principle, nontrivial.
In this paper, a more comprehensive analysis is presented.
In particular, a technique is proposed, independent on both the
nuclear potential and the nucleon model used in the calculation,
able to take into account the nuclear effects included in the IA analysis
and to safely disentangle the neutron information from them, 
even at moderate values of the
momentum transfer. 
Thanks to this observation, coherent DVCS off $^3$He
is strongly confirmed as a key experiment 
to access the neutron GPDs.

The paper is structured as follows. In the next section, 
part of the formalism used
in the IA analysis, only sketched in the previous
Rapid Communication, Ref. \cite{nostro}, is developed and motivated. 
In the third one, 
the ingredients used in the calculation are described. 
In the following section, the proton
and neutron contributions to the $^3$He GPDs, obtained in the present approach,
are discussed, together with their integrals, providing correctly 
the electromagnetic form factors (ffs). In the fifth section, 
it comes the most relevant result
of the paper: a safe extraction procedure of the neutron
information from $^3$He data, independent on both the
nuclear potential and the nucleon model used in the calculation. 
The impact of this study in the present 
experimental scenario is eventually discussed
in the Conclusions.
Two Appendixes have been added, detailing the 
description of $^3$He GPDs in terms of few-body wave functions.

\section{Impulse Approximation Analysis of $^3$He GPDs}

Let us first remind the main properties of GPDS
(for comprehensive reviews, see, e.g., \cite{dpr,rag,bp}),
to really understand the importance of measuring the neutron GPDs
and the advantages offered by $^3$He. 

For a spin $1/2$ hadron target $A$, with initial (final)
momentum and helicity $P(P')$ and $s(s')$, 
respectively, 
the GPDs $H_q(x,\Delta^2,\xi)$ and
$E_q(x,\Delta^2,\xi)$
are introduced through the light cone correlator
\begin{eqnarray}
\label{eq1}
F_{s's} ^{q,A,\mu}\hspace{-0.55em}& ( & \hspace{-0.55em} x,\Delta^2,\xi) =
\int {\frac{d z^-}{4 \pi}} e^{i x \bar P^+ z^-}
{_A\bra} P' s' | 
\hat O_q^\mu
| P s \ket_A |_{z^+=0,z_\perp = 0} 
\nonumber \\
&=& \frac{1}{2 \bar P^+} 
\Big [ H_q^A(x,\Delta^2,\xi) \bar u(P',s') 
\gamma^\mu u(P,s) \Big.
\nonumber
\\
&+&
\Big . 
E_q^A(x,\Delta^2,\xi) \bar u(P',s') 
{i \sigma^{\mu \alpha} \Delta_\alpha \over 2m} u(P,s) \Big ]+...~,
\end{eqnarray}
where $\hat O_q^\mu=
\bar \psi_q 
\left(- {z \over 2 } \right)
\gamma^\mu \, 
\psi_q \left( {z \over 2 } \right)
$,
$\bar P=(P+P')/2$,
$\psi_q$ is the quark field, $m$ is the hadron mass and
$q^\mu=(q_0,\vec q)$.
Ellipses denote higher twist structures.
The $\xi$ variable, the so-called
skewedness, 
is 
$
\xi = - {\Delta^+ / (2 \bar P^+)}
$
(everywhere in this paper, $a^{\pm}=(a^0 \pm a^3)/\sqrt{2}$).
In addition to the variables
$x,\xi$ and $\Delta^2$, GPDs depend
on the momentum scale $Q^2$. 
Such a dependence, irrelevant in this investigation, 
is not shown in the following.

In proper limits, GPDs are related to known quantities,
such as the PDFs and the electromagnetic
form factors. In particular, the following 
constraints, relevant for what follows, hold:

i) in the 
``forward'' limit, 
$P^\prime=P$, where $\Delta^2=\xi=0$, 
DIS Physics is recovered, and
$H_q^A(x,\Delta^2,\xi)$ coincides with the usual PDF,
$
H_q^A(x,0,0)=q(x)
$,
while $E_q^A(x,0,0)$ {\it is not accessible};

ii)
the integration over $x$ gives, for
$H_q$ ($E_q$),
the contribution
of the quark of flavour $q$ to the Dirac (Pauli)
ff of the target:
\bq
\label{gpdff}
\int_{-1}^1 
dx\, H_q^A(E_q^A)(x,\Delta^2,\xi) = F_{1(2)}^{A,q}(\Delta^2)~.
\eq

iii) the polynomiality property,
involving higher moments of GPDs, according to which
the $x$-integrals of $x^nH^q$ and of $x^nE^q$
are polynomials in $\xi$ of order $n+1$.

For later convenience, let us define the following auxiliary
function, given simply by the sum of the GPDs $H_q^A$ and $E_q^A$
for a given target $A$ of spin $1/2$:

\begin{equation}
\label{gmgpd}
\tilde G_M^{A,q}(x,\Delta^2,\xi) = 
H_q^A(x,\Delta^2,\xi)+E_q^A(x,\Delta^2,\xi)~.
\end{equation}

This function, due to Eq. (\ref{gpdff}), fulfills obviously the
following relation

\bq
\label{gmff}
\int_{-1}^1 
dx\, \tilde G_M^{A,q}(x,\Delta^2,\xi) & = & 
F_1^{A,q}(\Delta^2) 
+ F_2^{A,q}(\Delta^2)
\nonumber
\\
& \equiv & G_M^{A,q}(\Delta^2)~,
\eq
being $G_M^{A,q}(\Delta^2)$ the contribution
of the quark of flavour $q$ to the magnetic ff of the target $A$.
 
A fundamental result is Ji's sum rule (JSR)
\cite{Ji:1996ek}, according to which the forward limit
of the second moment of the unpolarized GPDs is related
to the component, along the quantization
axis, of the total angular momentum of the quark $q$
in the target $A$, $J_{q}^A$, according to 
\begin{eqnarray}
\label{sumji2}
J_{q}^A  = 
	\int_{-1}^1 
dx\, x  \,
\tilde G_M^{A,q}(x,0,0) ~.
\end{eqnarray}
The combination $\tilde G_M^{N,q}=H_q^N + E_q^N$ is 
therefore needed to study the angular momentum
content of the nucleon $N$, through the JSR, and 
OAM could be obtained from
$J_{q}^A$,
being the helicity content measurable in DIS
and SiDIS. The measurement of GPDs is therefore very helpful
for the understanding of the puzzling spin structure
of the nucleon. The proton data alone do not allow the 
flavor decomposition of the GPDs and therefore,
as it happens for any other parton observable, 
the neutron data are very important.
{
To obtain the latter, in principle, among the light nuclei,
$^3$He is an ideal target. 
To understand this fact it is sufficient to realize that,
at low $\Delta^2$, the GPDs $E$ of protons and neutrons have
similar size and opposite sign (their first moments are
$\kappa_p$=1.79 $\mu_N$ and $\kappa_n$=-1.91 $\mu_N$, respectively,
being $\mu_N$ the nuclear magneton). This makes
any isoscalar nuclear target, such as $^2$H or $^4$He, not suitable
for the extraction of the neutron $E$, basically canceled by the proton one,
in the coherent channel of DVCS. As a matter of facts, the contribution
of the proton and neutron GPD $E$ to the deuteron GPDs has been
neglected in the IA calculation presented in Ref. \cite{cano1}.
Accordingly, in the recent analysis in Ref. \cite{liuti4}, the contribution
of the nucleonic GPD $E$ to the angular momentum carried by the quark 
in the deuteron
has been found to be very small, if only conventional nuclear
effects are taken into account.

On the contrary, in the $^3$He case, 
GPDs are found to be sensitive to the nucleon $E_q$.
One should in fact notice that $\mu_3 \simeq -2.13 \mu_N$,
a value rather close to the neutron one, $\mu_n \simeq -1.91 \mu_N$.}
As it is well known, $\mu_3$ and $\mu_n$ would be equal,
i.e., there would be no proton contribution to $\mu_3$, if $^3$He could be
described by an independent particle model with central
forces only. Of course this scenario is a rough approximation; nonetheless,
realistic calculations show that the system lies in this configuration
with a probability close to 90 \% \cite{friar}, allowing
a safe extraction of the neutron DIS structure functions 
from $^3$He data, as suggested in \cite{antico,SS},
estimating effectively nuclear corrections using static properties.
Here, the situation is in principle somehow different, 
because GPDs are not densities. 
Anyway, this scenario is recovered at least
in the forward limit, where the JSR holds:
static $^3$He properties can be again advocated.

In a previous Rapid Communication of ours, Ref. \cite{nostro},
it has been established to what extent, 
close to the forward limit and slightly beyond it, 
the measured GPDs of $^3$He are dominated by the neutron ones.
Ref. \cite{nostro} represents
a pre-requisite for any experiment of
coherent DVCS  off $^3$He, an issue which is under 
consideration at JLab.
{In the rest of this section, part of the formalism used
in the IA analysis, only sketched in the previous
Rapid Communication, Ref. \cite{nostro}, is summarized}.

The GPD 
$H_q^3$ of $^3$He has been evaluated, in IA, already in
Ref. \cite{io}.
Let us see now how the scheme can be generalized to
obtain also the combination of GPDs $\tilde G_M^{3,q}$, Eq. (\ref{gmgpd}).
First of all, one should realize that,
in addition to the kinematical variables
$x$ and $\xi$, one needs the corresponding ones 
for the nucleons in the target nuclei,
$x'$ and $\xi'$. 
These quantities can be 
obtained introducing the ``+''
components of the momentum $k$ and $k + \Delta$ of the struck parton
before and after the interaction, with respect to
$\bar p^+ = {1 \over 2} (p + p')^+$,
being $p(p')$ the initial (final) momentum of the
interacting bound nucleon  \cite{io}:
\begin{eqnarray}
\xi' & = & - { \Delta^+ \over 2 \bar p^+}~,
\label{defxi1}
\\
x' & = & {\xi' \over \xi} x~,
\label{defx1}
\end{eqnarray}
and, since $\xi = - \Delta^+ / (2 \bar P^+)$, one has
\begin{eqnarray}
\xi ' = {\xi \over  {p^+ \over P^+}( 1 + \xi ) - \xi}~.
\label{xi1}
\end{eqnarray}

Now, the standard procedure developed in IA studies of
DIS off nuclei (see, i.e., \cite{fs}) is applied to obtain,
for $H_q^3,\,\tilde G_M^{3,q}$, 
convolution-like equations
in terms of the corresponding
nucleon quantities, $H_q^N,\,\tilde G_M^{N,q}$.
Let us just recall the main steps of the derivation.
First of all, since the nuclear states have to
be described by non-relativistic (NR) wave functions, the nuclear
overlaps and the one-body operator $\hat O_q^\mu$ in Eq. (\ref{eq1})
are treated in a NR manner; then,  proper components
of the $\gamma^\mu$ matrix and proper combinations of 
the nuclear and nucleon spin projections 
are selected to extract, from the NR reduction
of the correlator Eq. (\ref{eq1}),
independent relations for $H_q^3,\,\tilde G_M^{3,q}$.
This procedure is detailed in Appendix A.
The final result is, for any spin 1/2 target $A$ (cf. Eq. (\ref{eq1})):
\begin{eqnarray}
\label{relazioni}
H_q^A(x,\Delta^2,\xi) &=& \dfrac{\overline{P}^+}{m} F_{++}^{q,A,0}
(x,\Delta^2,\xi)~,
\nonumber \\
\tilde G_M^{A,q}(x,\Delta^2,\xi) &=& \dfrac{2\overline{P}^+}{\Delta_z}
F_{+-}^{q,A,1}(x,\Delta^2,\xi)~.
\end{eqnarray}
Once the correct NR treatment is identified, IA is applied.
The detailed machinery is thoroughly explained in Ref. \cite{io}
for the case of $H_q^3$ and it is not repeated here.
The $\tilde G_M^{3,q}$ result, already presented in Ref.
\cite{nostro}, is obtained analogously.
The procedure requires
the insertion of complete sets of states to the left
and to the right hand side
of the one-body operator $\hat O_q$ in Eq. (\ref{eq1}),
so that one-body matrix elements and nuclear overlaps are
identified thanks to the IA.
Using then Eq. (\ref{relazioni}) to obtain relations
between nuclear and nucleon GPDs, from the ones
between nuclear and nucleon correlators,
the following convolution-like formulae are eventually found:

\bq
\label{H}
H_{q}^{3}(x,\Delta^2,\xi) &=& 
\sum_N
\int dE 
\int d\vec{p}
\,
\overline{\sum_{ S }}
\sum_{s}
P^N_{SS,ss}(\vec p,\vec p\,',E)  
\nonumber \\
& \times &
{\xi' \over \xi}
H^{N}_q(x',\Delta^2,\xi')~,
\eq
\bq
\label{HpE}
\tilde G_M^{3,q}(x,\Delta^2,\xi) & = &  
\sum_N
\int dE 
\int d\vec{p}
\nonumber \\
& \times &
\left [ P^N_{+-,+-}(\vec p,\vec p\,',E)  
-
P^N_{+-,-+}(\vec p,\vec p\,',E) \right ]
\nonumber \\
& \times &  
{\xi' \over \xi}
\tilde G_M^{N,q}(x',\Delta^2,\xi')~.
\eq
In Eqs. 
(\ref{H}) and (\ref{HpE}), 
proper components
appear of 
$P^N_{SS',ss'}(\vec p,\vec p\,',E)$,
the spin-dependent
non-diagonal spectral function
of the nucleon $N=p,n$ in $^3$He, being
$S,S'(s,s')$ the nuclear (nucleon) spin projections
in the initial and final state, respectively,
and $E= E_{min} +E_R^*$, 
being $E^*_R$ the excitation energy 
of the two-body recoiling system
and 
$E_{min}=| E_{^3He}| - | E_{^2H}| = 5.5$ MeV. 
To calculate the spectral function,
one has to evaluate intrinsic overlap integrals.
The relations between $P^N_{SS',ss'}(\vec p,\vec p\,',E)$,
the overlaps and the
2- and 3-body radial wave functions at our disposal are
discussed and explained in the Appendix B.

\begin{figure}[t]
\includegraphics[height=6cm]{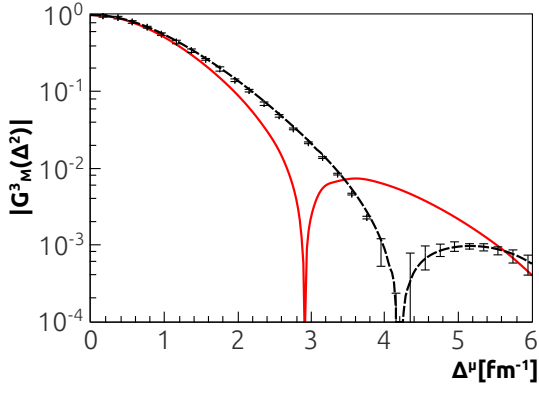}
\caption {(Color online)
The magnetic ff of $^3$He, $G_M^3(\Delta^2)$.
Full line: the present IA calculation, obtained as the integral of
Eq. (\ref{HpE}), summed over the
flavors (see text), together with the experimental data
\cite{dataff}. The units used for the momentum transfer,
$\Delta^\mu \equiv \sqrt{- \Delta^2}$,
have been chosen for an easy comparison with the results of Ref.
\cite{Marcucci:1998tb}.
}
\end{figure}

\begin{figure}[t]
\includegraphics[height=6cm]{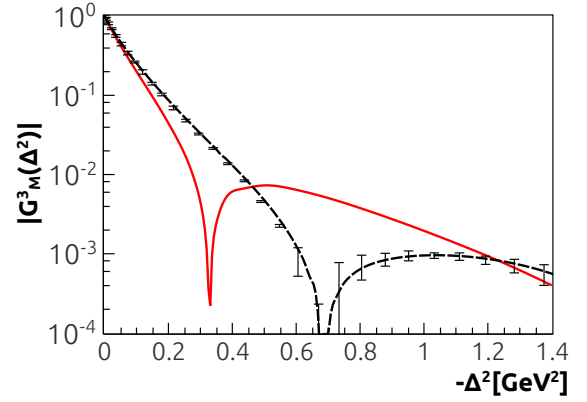}
\caption {(Color online)
As in Fig. 1, but in units of GeV$^2$,
as a function
of $- \Delta^2$.
}
\end{figure}

As discussed in Ref. \cite{io},
the accuracy of this calculation,
since a NR spectral function
is used to evaluate Eqs. 
(\ref{H}) and (\ref{HpE}),
is
of order 
${\cal{O}} 
\left ( {\vec p\,^2 / M^2},{\vec \Delta^2 / M^2} \right )$.
The interest of the present
calculation is precisely that of investigating nuclear effects at low values
of $\vec \Delta^2$, for which measurements in the coherent channel 
may be performed.

The calculation of Eq. (\ref{H}) has been performed
and discussed in Ref. \cite{io}. It has been found that the
properties $i), ii)$ of GPDs, defined above, are properly
recovered. The property $iii)$, the so called polinomiality, 
is slightly violated at the low values of $\Delta^2$ and $\xi$
of interest here, due to the NR IA which is being used.
The same problem affects also the evaluation of Eq. (\ref{HpE}).

In order to evaluate Eq. (\ref{HpE}), the quantity
of interest here,
one needs models of the nuclear spectral function
and of the nucleon GPDs $H_q^N$ and $E_q^N$.
The models used in the present calculation 
are described in the next section.

\section{Ingredients of the calculation}

{As stated above and as it is explained thoroughly
in the Appendix B, in order to evaluate
$P^N_{SS',ss'}(\vec p,\vec p\,',E)$,
the spin-dependent
non-diagonal spectral function
of the nucleon $N$ in $^3$He,
one needs intrinsic nuclear overlaps.
These quantities have been evaluated exactly 
in Ref. \cite{pssk}
along the line of Ref. \cite{gema}, 
using the wave function of Ref. \cite{tre}
corresponding to the Av18 interaction \cite{av18},
taking into account the Coulomb repulsion between
the two protons.
For a relevant comparison to be described later on,
the calculation has been performed also using 
the nuclear overlaps corresponding to the Av14 interaction
\cite{av14}, 
firstly evaluated to be used in Ref. \cite{gema}, 
including again the Coulomb repulsion.}

{Concerning the model of the nucleon GPDs to be used,
a preliminary observation is in order. In this paper the interest
is only in the evaluation
of nuclear effects, to test if the neutron dominates the nuclear observable 
and to suggest an extraction procedure. It is therefore important to use
different models, based on different assumptions on the hadron structure,
to evaluate $\tilde G_M^3$, and to check if the dominance of the neutron
contribution, and the reliability of the extraction procedure, are valid
independently of the model used. We are going therefore to use
three very different models, namely:

1) the model of Ref. \cite{rad1}, which, despite of its
simplicity, fulfills the general properties of GPDs.
The GPD $H_q$ is built in agreement with
the Double Distribution representation \cite{radd}.
The $\Delta^2$ dependence, factorized out from the one
in $x$ and $\xi$, is given by
$F_q(\Delta^2)$, i.e., the contribution
of the quark of flavour $q$
to the nucleon ff, obtained from
the experimental values of the proton, $F_1^p$, and
of the neutron, $F_1^n$, Dirac ffs (see Ref. \cite{io}).
For the numerical evaluation,
use has been made of the parametrization of the nucleon
Dirac ff given in Ref. \cite{gari}.
The model has been minimally extended to parametrize also the
GPD $E_q$, assuming that
it is proportional to the 
charge of $q$ (this choice, a very natural one, is used, 
e.g., in Refs. \cite{from}).
In this way, proper relations are found between
the $u$ and $d$ contributions to the Pauli ff and the
proton and neutron Pauli ff; again, 
for the latter quantities, the parametrization of Ref. \cite{gari}
has been used;

2) The model of GPDs
described in Ref. \cite{sv}, arising in a constituent quark
description of the nucleon structure, 
performing a microscopic calculation
without assuming any factorization.
The model, completely different
in spirit from the previously described one, 
refers to a low $Q^2$ scale and
is reasonable only in the valence quark region;

3) The microscopic model calculation
of Ref. \cite{song}, based on a simple version of the
MIT bag model, i.e., assuming 
confined free relativistic quarks
in the nucleon, a completely different 
scenario with respect to the constituent
quark picture and the double distribution one.

For the present aim, there is no need 
to use recent, sophisticated GPDs models,
such as the ones of Refs. \cite{belli1,belli2},
which would complicate the description without 
adding new insights.
Once the experiments are planned, more realistic calculations
involving phenomenologically motivated models will be easily implemented
in our scheme. }

\section{The neutron and proton contributions to the GPDs of $^3$He}

Using the ingredients described above,
the calculation of Eq. (\ref{HpE})
has been performed in the nuclear Breit-frame.
Unfortunately,
the only safe way to 
establish the reliability of the approach,
the comparison with experiments, is not possible.  
In facts, data for the GPDs are not available and, for $E_q^3$
in particular, even the forward limit is unknown.
One check is in any case possible and it is therefore a crucial one: 
the quantity 
\bq
\label{sum}
\tilde G_M^{3}(x,\Delta^2,\xi) = \sum_q \tilde G_M^{3,q}(x,\Delta^2,\xi), 
\eq
i.e., Eq. (\ref{gmgpd}), summed over the active flavors,
can be integrated over $x$ to give the 
experimentally well-known magnetic ff of $^3$He, 
$G_M^3(\Delta^2)$ (cf. Eq. (\ref{gmff})).
The result obtained using this procedure
is in excellent agreement with the Av18
one-body calculation presented in Ref. \cite{Marcucci:1998tb}, and with
the non-relativistic part of the calculation in Ref. 
\cite{Baroncini:2008eu} (see Figs. 1-2).
Moreover, for the values of $\Delta^2$ which are relevant for the 
coherent process under investigation here,
i.e., $-\Delta^2 \lesssim 0.15$ GeV$^2$,
our results compare well also with the data
\cite{dataff}.
For higher values, the agreement is lost and
to get a good description one should go beyond IA, including three-body forces 
and two-body currents
(see, e.g., \cite{Marcucci:1998tb}).
If measurements were performed at {these} values of $-\Delta^2$,
our calculations could be improved by allowing for these effects;
for the moment being,
since coherent DVCS
cannot be measured at $-\Delta^2 >$ 0.15 GeV$^2$ for nuclear targets,
the good description obtained close to the static point is quite
satisfactory.

\begin{figure}[t]
\includegraphics[height=6cm]{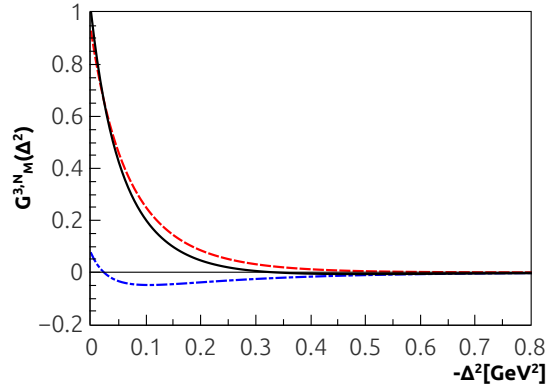}
\caption {(Color online)
The proton contribution, $G_M^{3,p}(\Delta^2)$ (dot-dashed),
and the neutron one, $G_M^{3,n}(\Delta^2)$ (dashed), to
the magnetic ff of $^3$He, $G_M^3(\Delta^2)$ (full),
obtained within the present
IA approach.
}
\end{figure}

\begin{figure}[t]
\includegraphics[height=6cm]{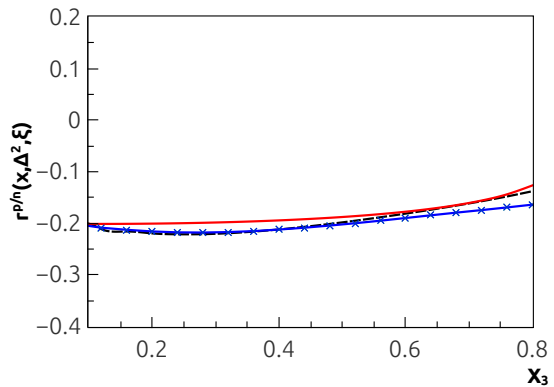}
\caption {(Color online)
The ratio of the proton to neutron contribution
to the quantity $x_3 \tilde G_M^{3}(x,\Delta^2,\xi)$ 
of $^3$He (full), 
Eq. (\ref{rat_pn}),
obtained at $-\Delta^2 =
0.1$ GeV$^2$ and $\xi_3$ = 0,
using the nucleon GPDs model of Ref. \cite{rad1}
in Eq. (\ref{HpE}),
compared with the same ratio evaluated using the
nucleon GPDs model of Ref. \cite{sv} (dashed)
and that of Ref. \cite{song} (crosses). 
}
\end{figure}

With the comfort of this successful check,
let us briefly recall the main outcome of Ref. \cite{nostro}.
{
In that paper, the quantity $x_3\tilde G_M^{3}(x,\Delta^2,\xi)$, which,
in the forward limit, yields
the integrand of the JSR
(cf. Eq. (\ref{sumji2}) where the relation is given
for a given flavor $q$), has been presented as a function of $x_3$,
in the forward limit and at finite values of
$\Delta^2$ 
and $\xi_3$
(Here we have defined
$x_3 = ({M_A/ M})x$ and $\xi_3=({M_A/ M}) \xi$, in order to
recover the standard notation used in studies of DIS off nuclei.
In facts, the Bjorken variable 
is defined as $x_{Bj}=Q^2/(2 M \nu)$, being $\nu=q_0$ in the laboratory 
system. It ranges naturally between 0 and ${M_A/ M}$
for a nuclear target of mass $M_A$.
It is therefore convenient to rescale the variables
$x,\xi$ by the factor ${M_A/ M}$).

For the nuclear GPDs, a dramatic $\Delta^2$ behavior,
basically governed by that of the ff, has been found. 
This fact can be realized looking at Fig. 3, where the separate contribution
of the neutron and of the protons to $G_M^{3}(\Delta^2)$
is shown in linear scale.
The most striking result of Ref. \cite{nostro} is actually}
that the contribution of the neutron is impressively
dominating the nuclear GPD at low $\Delta^2$, with the proton
contribution growing fast with increasing $\Delta^2$.
{However for the flavor $d$ the impressive
dominance of the neutron contribution varies slowly with increasing 
$\Delta^2$. 
This behavior of the $d$ contribution
is explained in Ref. \cite{nostro} in terms of the flavor structure of Eqs.
(\ref{H}) and (\ref{HpE}).}

Of course, the shape of the curves obtained for
$\tilde G_M^3$
is very dependent 
on the nucleonic model 
used as input in the calculation, i.e.,
in the case of Ref. \cite{nostro}, that of Ref. \cite{rad1}. 
One should not forget, {we reiterate}, that
the aim of this analysis, for the moment being,
is that of getting a clear estimate of the 
proton and neutron contributions to the nuclear observable,
a feature rather independent on the nucleonic model.
To demonstrate this property, we have plotted, in
Fig. 4, the ratio of the proton to neutron contribution to
$\tilde G_M^3$:

\begin{equation}
\label{rat_pn}
r^{p/n}(x,\Delta^2,\xi) = \tilde G_M^{3,p}(x,\Delta^2,\xi)/
\tilde G_M^{3,n}(x,\Delta^2,\xi)~,
\end{equation}
calculated using as input the $\tilde G^{p(n),q}_M$ corresponding
to the model of Ref. \cite{rad1}, and those of 
the very different models of Refs. \cite{sv} and \cite{song}. 
The {three} ratios are slowly varying as a function of $x_3$ and 
are very close to each other, demonstrating the very weak model
dependence of this feature of the result.

Summarizing, our IA calculation 
shows that, at very low $-\Delta^2 \simeq 0.1$ 
GeV$^2$, there is a clear dominance of the neutron contribution on 
the proton one, and that such a dominance, even stronger for the $d$ flavor,
does not depend on the nucleonic model used in the calculation.
Anyway, a couple of items have still to be investigated:
$i)$ even if the neutron contribution is dominating, the extraction
of the neutron GPDs from it may be nontrivial and a proper
strategy has to be studied;
$ii)$ although the very low $\Delta^2$ values define the most
interesting region, where, for example, the JSR can be checked,
this region may be difficult to be reached experimentally.
A procedure able to take into account the nuclear effects
arising in an IA description also at higher
$\Delta^2$ would be very helpful.

A positive answer to these two remaining problems will be given
in the next section. 

\section{Extracting the neutron information from $^3$He data}

Let us discuss the most relevant issue.

It is convenient to rewrite our main equation, Eq. (\ref{HpE}),
in a different form,
defining a variable $z$ as follows:  
\begin{eqnarray}
z - {M_A \over M}
{ \xi \over \xi'} & = & z -  {M_A \over M}
\left [ {p^+ \over P^+} ( 1 + \xi  ) - \xi \right ]
\nonumber
\\
& = & z + \xi_3 -  {M_A \over M}{ p^+ \over P^+} ( 1 + \xi )
\nonumber
\\
& = & z + \xi_3  -  {M_A \over M}{ p^+ \over \bar P^+}~,
\end{eqnarray}
where 
use has been made of Eqs.
(\ref{defxi1}) -- (\ref{xi1}).
Eq. (\ref{HpE}) can be written therefore in the form
\begin{eqnarray}
\tilde G_M^{3,q}(x,\Delta^2,\xi) & = &  
\sum_N \int_{x_3}^{M_A / M} { dz \over z}
g_N^3(z, \Delta^2, \xi ) 
\nonumber
\\
& \times &
\tilde G_M^{N,q} \left( {x \over z},
\Delta^2,
{\xi \over z},
\right)~,
\label{main}
\end{eqnarray}
where the off-diagonal spin-dependent light cone momentum distribution
\begin{eqnarray}
g_N^3(z, \Delta^2, \xi ) & = &  
\int d E
\int d \vec p
\, \tilde P_N^3(\vec p, \vec p + \vec \Delta, E) 
\nonumber
\\
& \times &
\delta \left( z + \xi_3  - {M_A \over M} { p^+ \over \bar P^+ } \right)
\label{gq0}
\end{eqnarray}
has been introduced and the quantity
\begin{eqnarray}
\label{sdoffsp}
\tilde P_N^3(\vec p, \vec p + \vec \Delta,E) 
& = &
P^N_{+-,+-}(\vec p,\vec p\,',E)  
\nonumber 
\\
& - &
P^N_{+-,-+}(\vec p,\vec p\,',E)~,
\end{eqnarray}
for the sake of clarity, has been introduced.

\begin{table}
\begin{center}
\begin{tabular}{|c|c|c|c|c|}
\hline
$\Delta^2$ & $G^{3,p,point}_M$ &$G^{3,p,point}_M$ & 
$G^{3,n,point}_M$ & $G^{3,n,point}_M$ 
\\
$[$GeV$^2$$]$ & Av18 & Av14 & Av18 & Av14
\\
\hline
0 & -0.044 & -0.049 & 0.879 & 0.874\\
\hline
-0.1 & 0.040 & 0.038 & 0.305 & 0.297\\
\hline
-0.2 & 0.036 & 0.035 & 0.125 & 0.119 \\
\hline
\end{tabular}
\end{center}
\caption{ The proton and neutron contributions
to the magnetic point like ff of $^3$He, corresponding to the
Av18 \cite{av18} and Av14 \cite{av14} nuclear interactions, 
for three low values of $-\Delta^2$.}
\end{table}

Let us now perform the $x-$integral of the function
$\tilde G_M^{3,q}(x,\Delta^2,\xi)$, given by Eq. (\ref{main}).
One obtains:
\begin{eqnarray}
\int & dx & \, \tilde G_M^{3,q}  (x,\Delta^2,\xi) 
=
\nonumber 
\\
& = &
\sum_N
\int dx \int {dz \over z} g_N^3(z,\Delta^2,\xi)
\tilde G_M^{N,q} \left ( { x \over z}, \Delta^2, {\xi \over z} \right ) =
\nonumber
\\
& = &
\sum_N
\int d x' \tilde G_M^{N,q} (x',\Delta^2,\xi') \int d z 
g_N^3(z,\Delta^2,\xi) =
\nonumber
\\
& = &
\sum_N G_M^{N,q}  (\Delta^2)
\, G_M^{3,N,point} (\Delta^2)
\nonumber
\\
& = & G_M^{3,q} (\Delta^2)~.
\label{ffc}
\eq
\begin{figure}[t]
\includegraphics[height=6cm]{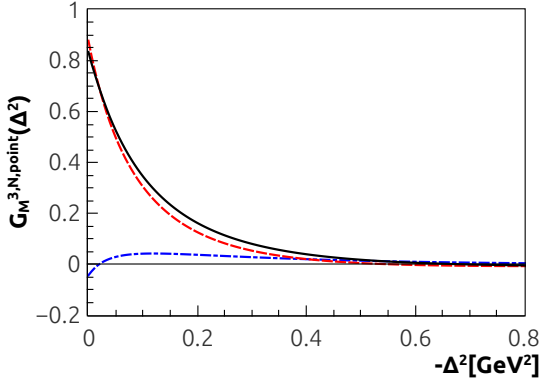}
\caption {(Color online)
The magnetic point like ff of $^3$He (full),
given
by the sum of the neutron (dashed) and proton (dot-dashed)
contribution (Eq. (\ref{pointlike})).
}
\end{figure}
In the equation above,
$
G_M^{3,q}
(\Delta^2)$ is the
contribution, 
of the quark of flavor $q$,
to the
nuclear ff;
$
G_M^{N,q}
(\Delta^2)$ is the contribution,
of the quark of flavour $q$,  
to ff of the nucleon $N$;
$
G_M^{3,N,point}
(\Delta^2)$ is the 
so-called $^3$He magnetic ``point like ff'', which
would represent the contribution of the nucleon $N$ to the
magnetic ff of $^3$He if $N$ were point-like.
The latter quantity has a relevant role
in our discussion; we stress that
it is given, in the present IA approach, by
\bq
\label{pointlike}
G_M^{3,N,point}(\Delta^2)
& = & \int dE \int d \vec p
\, \tilde P_N^3(\vec p, \vec p + \vec \Delta, E)
\nonumber
\\
& = & \int_{0}^{M_A / M} dz \, g_N^3(z,\Delta^2,\xi)~. 
\eq
This quantity, obtained in our Av18 framework, is shown for $N=p,n$,
and for their sum, $G_M^{3,point}(\Delta^2)$
in Fig. 5. Obviously, this one-body property can be obtained from 
the wave function only and,
at least for the low values of $\Delta^2$ of interest here, 
we checked that it depends very weakly
on the nuclear interaction.
In fact, we have performed the calculation using also the Av14 interaction,
including the Coulomb repulsion between the protons.
The curves we have obtained cannot be distinguished
from the Av18 ones shown in Fig. 5 and 
have not been reported.
The different magnetic point like ffs corresponding to the 
different nuclear interactions are therefore reported in Tab. 1
for three low values of $-\Delta^2$.
%
Now it comes an important observation.
The variable $z$, at the low values of $\Delta^2$ and $\xi$
which are of interest here, is very similar  
to the light cone $''+''$ momentum fraction
and the function $g_N^3(z,\Delta^2,\xi)$ 
very close to a standard, forward light-cone momentum
distribution. Besides,
the nucleon dynamics in $^3$He is, to a large
extent, a NR one. This makes any $^3$He light-cone momentum
distribution, polarized or unpolarized, strongly peaked
around $z=1$ (see, i.e., the discussion in Ref. \cite{io}).
Evaluating Eq. (24) close to $z=1$ means,
as a matter of facts, that all the nuclear
effects, but the ones due to the spin structure
of the target, are negligible. Basically, this
means that the momentum and energy distributions
of the nucleons do not affect the result.
If this is the case, from Eq. (\ref{main}) one gets:

\begin{eqnarray}
\tilde G_M^{3,q}(x,\Delta^2,\xi) & = &  
\sum_N 
\tilde G_M^{N,q} \left( x, \Delta^2, {\xi } \right)
\nonumber
\\
& \times &
\int_{x_3}^{M_A/M} 
\hskip -6mm 
{ dz }
g_N^3(z, \Delta^2, \xi ) + O(z-1)
\label{main1}
\end{eqnarray}
and, considering that, for $^3$He, 
$g_N^3(z, \Delta^2, \xi )\simeq 0$
for $x_3 \lesssim 1$, 
being this function strongly
peaked around $z=1$, the lower integration limit in $z$ can be put
equal to 0 for the $x_3$ values relevant here ($x_3 \lesssim 0.7$).
In other words, at low values of $\Delta^2$ and
$\xi$, the following approximation 
of Eq. (\ref{HpE}) should hold:
\begin{eqnarray}
\label{factorized}
\tilde G_M^{3,q}(x, \Delta^2,\xi) & \simeq & 
\sum_N 
\tilde G_M^{N,q} \left( x, \Delta^2, {\xi } \right)
\nonumber
\\
& \times &
\int_0^{M_A/M}{ dz }
g_N^3(z, \Delta^2, \xi ) 
\nonumber
\\
& = &
G^{3,p,point}_M(\Delta^2) \,
\tilde G_M^{p,q}(x, \Delta^2,\xi) 
\nonumber
\\
& + &
G^{3,n,point}_M(\Delta^2) \,
\tilde G_M^{n,q}(x,\Delta^2,\xi)~. 
\end{eqnarray}
Clearly, in the last line, use has been 
made of Eq. (\ref{pointlike}).
If the factorized expression above 
were a good approximation of Eq. (\ref{HpE}),
it would be very helpful.
In facts, Eq. (\ref{factorized}) is very simple: 
all the nuclear effects are hidden in the
magnetic point like ffs, quantities which can be obtained directly from
the wave function, without considering the complicated spectral
effects described by Eq. (\ref{HpE}); moreover, and more important,
these quantities, at low $-\Delta^2$, are theoretically well known
and, as Tab. 1 shows, depend very weakly on the
used nuclear interaction. 

Figs. 6 and 7 demonstrate that
Eq. (\ref{factorized}) approximates nicely the full result
Eq. (\ref{HpE}) in the kinematics of interest here, 
the two quantities differing of few percents at most 
for $x_3 < 0.7$. This is an important point:
Eq. (\ref{factorized}) can
now be solved to extract the neutron GPD
$\tilde G_M^{n}$:
\begin{eqnarray}
\label{extr}
\tilde G_M^{n,extr}(x, \Delta^2,\xi) & \simeq & 
{1 \over G^{3,n,point}_M(\Delta^2) }
\left\{ \tilde G_M^3(x, \Delta^2,\xi) \right.
\nonumber
\\
& - & 
\hskip -2mm
\left.
G^{3,p,point}_{M}(\Delta^2) 
\tilde G_M^p(x, \Delta^2,\xi) \right\},
\end{eqnarray}
an equation which could be used to obtain the neutron
information from $^3$He,  $\tilde G_M^3$,
and proton,
$\tilde G_M^p$, data, simply correcting by means
of the well known magnetic point like ffs.

\begin{figure}[t]
\includegraphics[height=6cm]{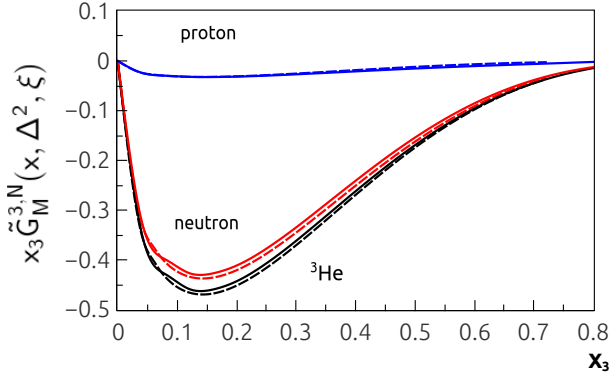}
\caption {(Color online)
The quantity 
$x_3 \tilde G_M^{3}(x,\Delta^2,\xi)$ 
of $^3$He in the forward limit, together with the proton and 
neutron contribution
(full lines), compared with the approximations to these quantities
given by the factorized form Eq. (\ref{factorized}) (dashed lines).
}
\end{figure}

\begin{figure}[t]
\includegraphics[height=6cm]{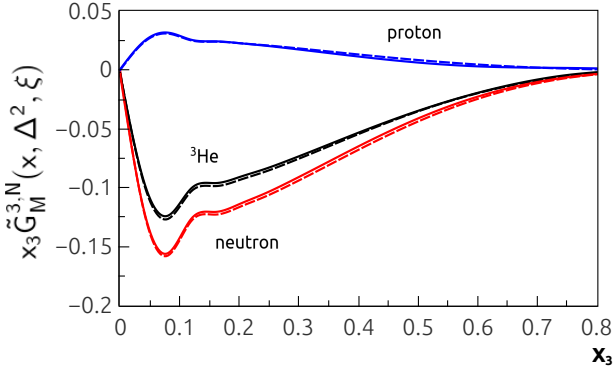}
\caption {(Color online)
The same as in Fig. 6, but at $\Delta^2 = -0.1$ GeV$^2$
and $\xi_3 = 0.1$.
}
\end{figure}

We have checked the validity of the proposed extraction procedure
by evaluating Eq. (\ref{extr}), using for the magnetic point like ffs
the ones corresponding the Av18 interaction, for
the proton GPD $\tilde G_M^p$ the one given by the model
\cite{rad1}, for $\tilde G_M^3$ the one calculated by means
of Eq. (\ref{HpE}), i.e., we are simulating $^3$He data
by using our best calculation.
If the extraction procedure were able 
to describe exactly, in an effective way, the nuclear
corrections predicted in IA, the obtained $\tilde G_M^{n,extr}$
should be equal to the neutron quantity, $\tilde G_M^{n}$,
used as input in the calculation,
i.e., again, the one predicted for the neutron by the model of Ref.
\cite{rad1}.

\begin{figure}[t]
\includegraphics[height=6cm]{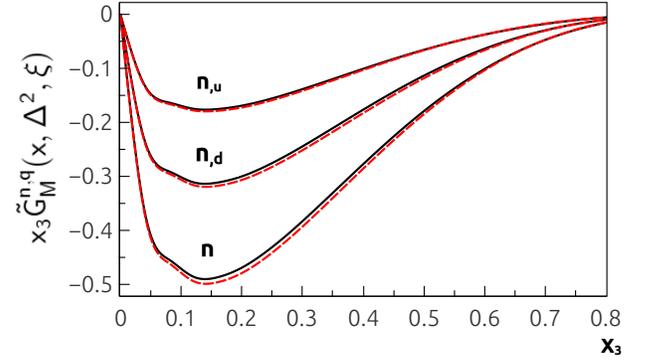}
\caption {(Color online)
The quantity 
$x_3 \tilde G_M^{n,q}(x,\Delta^2,\xi)$ 
for the neutron in the forward limit,
evaluated in the model of Ref. \cite{rad1}
and labelled $n$, given by the sum of
the $u$, labelled $n_u$, and $d$, labelled
$n_d$ contribution (full lines), compared 
with the approximations to these quantities given by Eq.
(\ref{extr}) (dashed lines) (see text).  
}
\end{figure}

\begin{figure}[t]
\includegraphics[height=6cm]{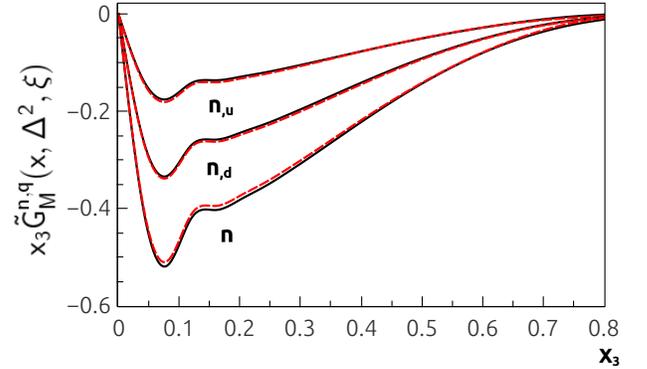}
\caption {(Color online)
The same as in Fig. 8, but at $\Delta^2 = -0.1$ GeV$^2$
and $\xi_3 = 0.1$. 
}
\end{figure}

Figs. 8 and 9 demonstrate that these
two quantities, and the $u$ and $d$ flavor contribution
to them, differ at most by few percents in all the
relevant kinematical range.
This means that the growth of the proton contribution to the $^3$He
observable, which seemed to hinder the extraction of the neutron
information, in particular for the $u$ flavor, is governed 
by the $-\Delta^2$ behavior of the magnetic point like ffs,
quantities which are under theoretical control.
By using them, the extraction of the neutron GPDs, close to the forward
limit, is safe and basically model independent.

\begin{figure}[t]
\includegraphics[height=6cm]{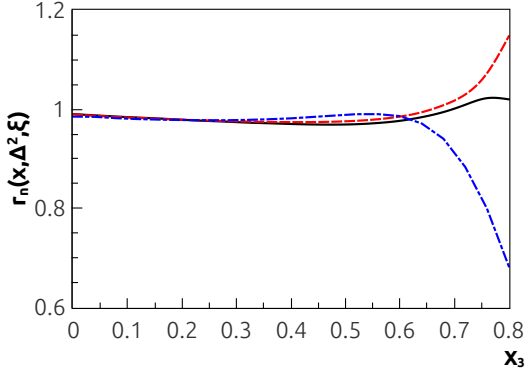}
\caption {(Color online)
The ratio Eq. (\ref{rn}), in the forward limit (full),
at $\Delta^2 = - 0.1$ GeV$^2$ and $\xi_3 = 0$ (dashed)
and 
at $\Delta^2 = - 0.1$ GeV$^2$ and $\xi_3 = 0.1$ (dot-dashed).
}
\end{figure}

\begin{figure}[t]
\includegraphics[height=6cm]{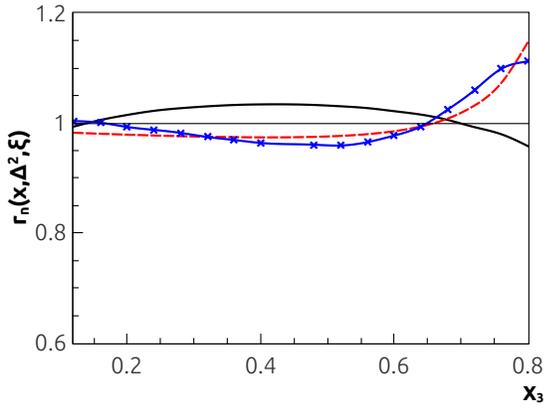}
\caption {(Color online)
The ratio Eq. (\ref{rn}),
at $\Delta^2 = 0.1$ GeV$^2$ and $\xi_3 = 0$,
using, to evaluate Eq. (\ref{HpE}),
for the nucleon GPDs,
the model of Ref. \cite{rad1} (dashed),
the one of Ref. \cite{sv} (full)
and that of Ref. \cite{song} (crosses).
}
\end{figure}

\begin{figure}[t]
\includegraphics[height=6cm]{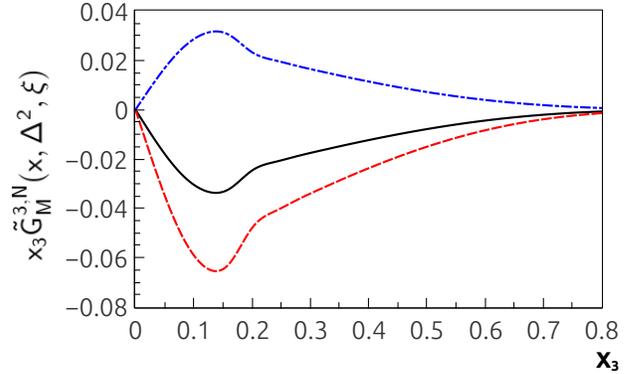}
\caption {(Color online)
The quantity $x_3 \tilde G_M^{3}(x,\Delta^2,\xi)$, 
at $\Delta^2 =-0.2$ GeV$^2$
and $\xi_3 = 0.2$ (full), together with the neutron (dashed)
and the proton (dot-dashed) contribution.
}
\end{figure}

\begin{figure}[t]
\includegraphics[height=6cm]{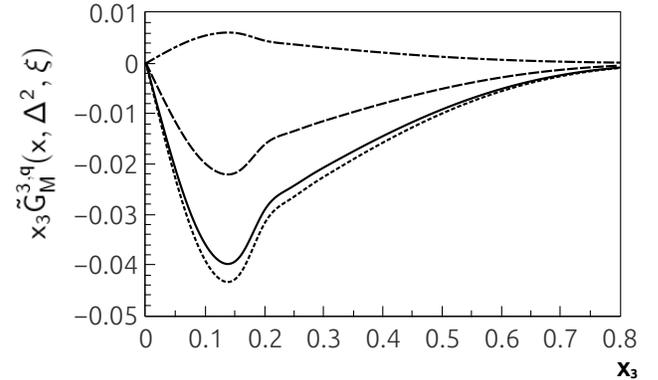}
\caption {
The quantity $x_3 \tilde G_M^{3,q}(x,\Delta^2,\xi)$  
for the $d$ (full) and $u$ (dot-dashed) flavor, at
$\Delta^2 = -0.2$ GeV$^2$,
and $\xi_3$=0.2. 
The neutron contributions for the $d$ (dashed) and $u$
(long-dashed) flavor are also shown.
}
\end{figure}

\begin{figure}[t]
\includegraphics[height=6cm]{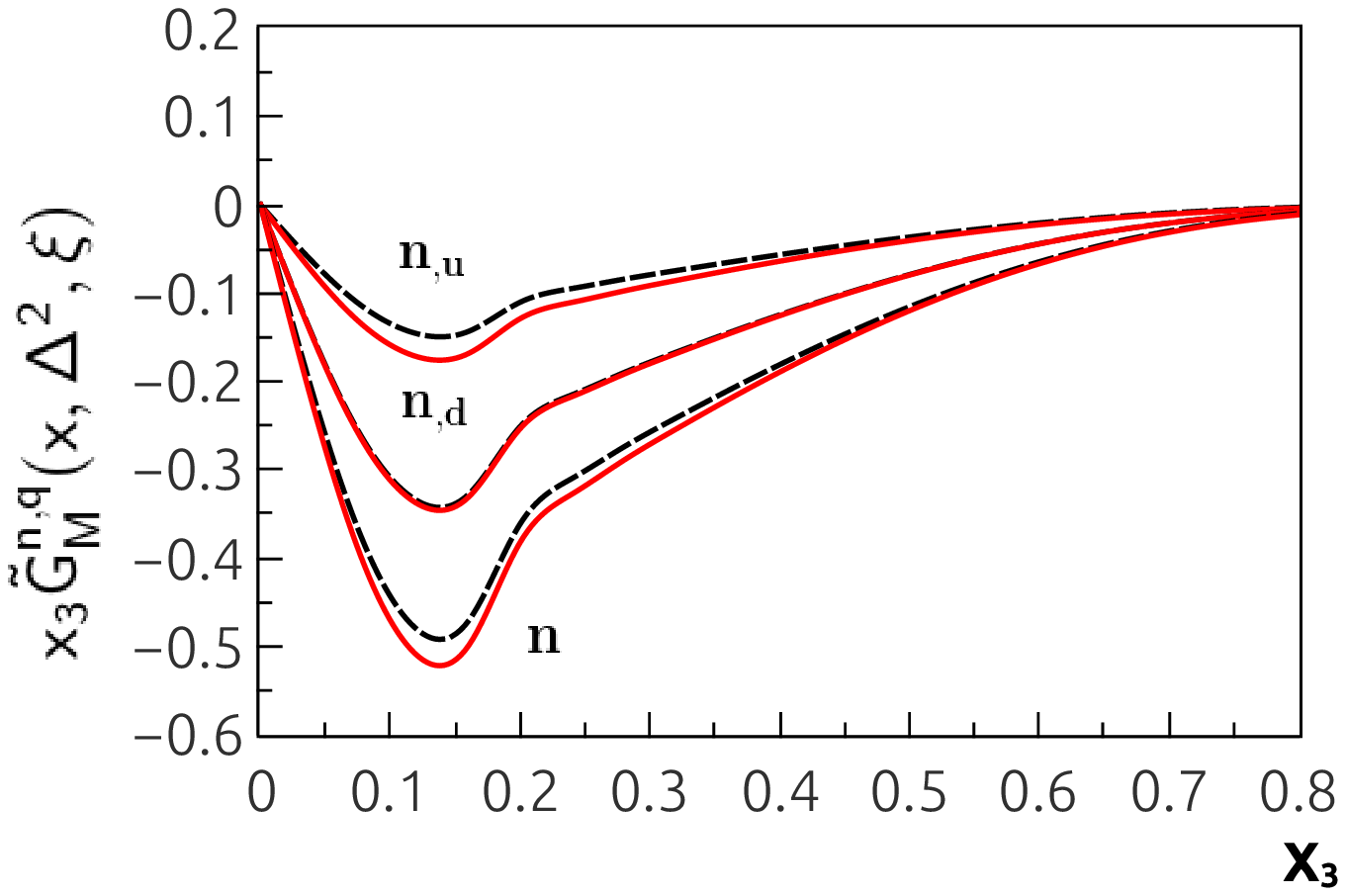}
\caption {(Color online)
The same as in Fig. 8, but at $\Delta^2 =-0.2$ GeV$^2$
and $\xi_3 = 0.2$.
}
\end{figure}

\begin{figure}[t]
\includegraphics[height=6cm]{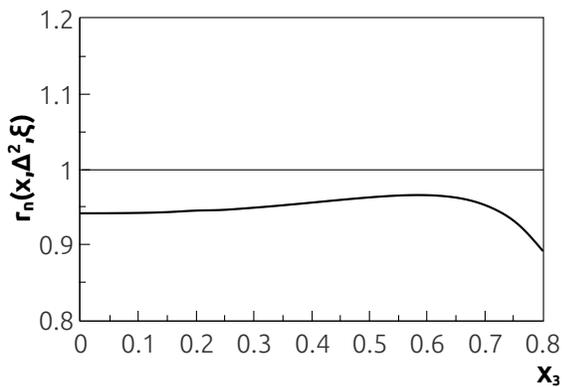}
\caption {
The same as in Fig. 10, but at $\Delta^2 =-0.2$ GeV$^2$
and $\xi_3 = 0.2$.
}
\end{figure}

These feature is even more evident in Fig. 10, where the ratio
\begin{eqnarray}
\label{rn}
r_n(x, \Delta^2,\xi) = \tilde G_M^{n,extr}(x, \Delta^2,\xi)
/
\tilde G_M^{n}(x, \Delta^2,\xi)~,
\end{eqnarray}
is shown. Obviously, this ratio would be one in case
the extraction procedure were working perfectly.
It is clear that, in all the relevant kinematics, the extraction
procedure takes into account the nuclear effects introduced
in the IA analysis for $x_3 \lesssim 0.7$ with an accuracy
of a few percents.

Besides, the quality of the extraction does not
depend on the nucleonic model used.
In Fig. 11, in one particular kinematics, it is shown the ratio
$r_n$, Eq. (\ref{rn}), obtained using the three different models
of GPDs considered in this work, i.e. the ones of Refs. \cite{rad1,sv,song}.
It is clearly seen that the procedure works 
within a few percent accuracy, independently on the model used.

Let us show now that the technique works at
higher $-\Delta^2$ values.
Figs. 12-15 are devoted to show this fact. All the steps are repeated,
in a sort of useful summary of the procedure.
In Fig. 12, the proton and neutron contributions
to $^3$He are shown at $\Delta^2=-0.2$ GeV$^2$
and $\xi_3 = 0.2$, and they are found to be comparable
in size; in Fig. 13, it is shown that this problem
is serious, in particular, for the
$u$ flavor. In Fig. 14 and 15 the good quality of the extraction
procedure is demonstrated even in this less forward situation,
by showing the extracted neutron GPD and its ratio to the
model used as input in the calculation, respectively.
This is a good piece of news for the experimental programme,
in case that extremely small values of $-\Delta^2$ could not be reached.
However, we reiterate that the
very good extraction scheme is really useful where IA 
provides a reasonable description of the process, a region
which coincides with  $-\Delta^2 \lesssim 0.15$ GeV$^2$.

{Another feature of the IA used here is the assumption that
only nucleonic degrees of freedom are considered in the analysis,
and that the nucleon structure is not modified by the nuclear medium.
The study of possible modifications, i.e. the study
of ``off-shell'' effects,
is certainly a relevant issue and it would deserve dedicated 
investigations, beyond the
scope of the present paper. We note however that careful analyses of these 
effects for different targets, performed, e.g., in Ref. \cite{liuti3}
for $^4$He, show that they grow with $\Delta^2$, being
minimal close to the forward limit, where we are proposing
the relevant measurement.}

In closing this section, let us mention
that another process to access the
neutron GPDs is {\it incoherent} DVCS off the neutron 
in nuclear targets,
i.e., the process with the struck neutron detected
in the final state, in coincidence
with the scattered electron and the produced
photon. An experiment of this type will be performed
at the 12 GeV program of JLab \cite{silvia} for a $^2$H target.
We plan therefore to investigate
also incoherent DVCS off the neutron in $^3$He,
although these kind of processes could be spoiled
by Final State Interactions of the detected neutron. 

\section{Conclusions }

In this work we have thoroughly described an IA calculation 
of the GPDs $H_q$ and $E_q$ of $^3$He. In proper limits,
the correct constraints are recovered.
Coherent DVCS off $^3$He at low $\Delta^2$ turns out to be strongly dominated
by the neutron contribution, in particular for the $d$ flavor,
from which the neutron information has to be extracted.
A procedure has been described 
to take into account the nuclear effects included in the IA analysis
and to safely disentangle the neutron information from them, 
even at moderate values of the
momentum transfer. 
The only theoretical nuclear ingredients are
the neutron and proton contributions to the magnetic point like form factors
of $^3$He, quantities which are under good theoretical control
at low momentum transfer and that encode correctly all the nuclear
effects described in an IA framework.
If high values of $\Delta^2$ were reached, the IA description
would not be reliable and two-body currents and three-body
forces would have to be included into the approach.
We have checked that the dominance of the neutron contribution
and the proposed extraction procedure do not depend on the model
of the nucleon GPD used in the calculation.
Our results confirm strongly  
coherent DVCS off $^3$He
at low momentum transfer
as a key experiment 
to access the neutron GPDs.
It will be very interesting and useful
to perform a Light-Front analysis of the process,
which already started in SiDIS \cite{alessio}, 
so to have, from the
beginning, a relativistic framework for the investigation.

\acknowledgments
It is a pleasure to thank L.P. Kaptari, G. Salm\`e and E. Voutier for
enlightening discussions and suggestions.

\onecolumngrid

\appendix\section{GPDs from the components of the light cone correlator}

In this work, 
both the nucleus, being the relevant momentum transfer low,
and the nucleon, whose dynamics is governed by the Schr\"odinger equation,
are treated non-relativistically (NR).
A NR expression of the spinors appearing in 
Eq. (\ref{eq1})
has therefore to be used, in order to find 
explicit relations
between  the correlator $F^{q,A,\mu}_{s's} $,
Eq. (\ref{eq1}),
and the GPDs. 

In the NR limit one has (see, e.g., \cite{ps}):
\begin{eqnarray}
\label{spinor}
u(p,s) =
\left( \begin{array}{c}
 	\sqrt{p \cdot \sigma} \chi_s \\ \sqrt{p \cdot \overline{\sigma}} 
\chi_s
 \end{array}   \right)
\underset{N.R} \longrightarrow \sqrt{m} \left( \begin{array}{c}
\left( 1-\dfrac{\overrightarrow{p}\overrightarrow{\sigma}}{2m}  \right) \chi_s 
\\ \left( 1+\dfrac{\overrightarrow{p}\overrightarrow{\sigma}}{2m}  \right) 
\chi_s
\end{array}  \right)~,
\end{eqnarray}
\noindent

and:

\begin{eqnarray}
 \overline{u}(p',s') \gamma^0 u(p,s) &\approx &  2m \chi_{s'}^{\dagger} 
\chi_s~, 
\nonumber \\
 \overline{u}(p',s') \gamma^i u(p,s) &\approx &  
2m \chi_{s'}^{\dagger} \left( \dfrac{-i}{2m} 
\varepsilon^{ijk}\Delta_j\sigma_k \right) \chi_s~,
\nonumber \\
 \overline{u}(p',s')\left( \dfrac{i}{2m} 
\sigma^{i \nu}\Delta_{\nu} \right)u(p,s) &\approx &  2m \chi_{s'}^{\dagger} 
\left( \dfrac{-i}{2m} \varepsilon^{ijk}\Delta_j\sigma_k \right) \chi_s~. 
\end{eqnarray}
\noindent

Properly choosing the  components of $\gamma^\mu$, 
it is possible to find
the most convenient relation between
$F_{s's}^{q,A,\mu}$ and the GPDs.
By taking for $\chi$ 
the eigenstates of $\sigma^3$, 
being $z$ the quantization axis, fixing $\mu = 0$,  
one obtains:

\begin{eqnarray}
\label{gamma0}
F_{++}^{q,A,0} (x,\Delta^2,\xi)&=& \dfrac{m}{2 \overline{p}^+} 
H_q^A(x,\Delta^2,\xi) 
\nonumber \\
 F_{--}^{q,A,0}(x,\Delta^2,\xi)&=& F_{++}^{q,A,0}(x,\Delta^2,\xi)~,
\nonumber \\
 F_{-+}^{q,A,0}(x,\Delta^2,\xi) &=& F_{+-}^{q,A,0}(x,\Delta^2,\xi) = 0~.
\end{eqnarray}

\noindent
In the same way, for $\mu=1$ one gets

\begin{eqnarray}
 \label{gamma1}
F_{++}^{q,A,1} (x,\Delta^2,\xi) &=& 0~,
\nonumber \\
F_{--}^{q,A,1} (x,\Delta^2,\xi) &=& F_{++}^{q,A,1}(x,\Delta^2,\xi)~,
\nonumber \\
F_{-+}^{q,A,1}(x,\Delta^2,\xi) &=& - F_{+-}^{q,A,1}(x,\Delta^2,\xi) 
= - \dfrac{\Delta_z}{2 \overline{p}^+}  \tilde G_M^{A,q}(x,\Delta^2,\xi)~.
\end{eqnarray}

\noindent
Collecting all the above results, one has

\begin{eqnarray}
\label{finale}
H_q^A(x,\Delta^2,\xi) 
&=& \dfrac{\overline{p}^+}{m} F_{++}^{q,A,0} (x,\Delta^2,\xi)~,
\nonumber \\
\tilde G_M^{A,q}(x,\Delta^2,\xi) &=& \dfrac{2\overline{p}^+}{\Delta_z}
F_{-+}^{q,A,1}(x,\Delta^2,\xi)~,
\end{eqnarray}

which coincides with Eq. (\ref{relazioni}).



\section{The light cone correlator from nuclear overlaps in IA}

{The relevant nuclear structure quantity in the present calculation is
the spin-dependent non-diagonal spectral function of the nucleon $N$
in $^3$He: 

\begin{eqnarray}
 \label{spectral1}
 P^N_{SS',ss'}(\vec p,\vec p\,',E) 
&=& 
\dfrac{1}{(2 \pi)^6} 
\dfrac{M\sqrt{ME}}{2} 
\int d\Omega _t
\\
& \times &  
\sum_{\substack{s_t}} \langle\vec{P'}\,S' | 
\vec{p}\,' s',\vec{t}\,s_t\rangle_N
\langle \vec{p}\,s,\vec{t}\,s_t|\vec {P}\,S\rangle_N~,
\nonumber 
\end{eqnarray}
and the most important ingredient appearing in the definition
Eq. (\ref{spectral1}) is
the intrinsic overlap integral
\bq
\langle \vec{p}\,s,\vec{t}\,s_t|\vec {P}\,S\rangle_N
=
\int d \vec y \, e^{i \vec p \cdot \vec y}
\bra \chi^{s}_N,
\Psi_t^{s_t}(\vec x) | \Psi_3^S(\vec x, \vec y) \ket~
\label{trueover}
\eq 
between the wave function
of $^3$He,
$\Psi_3^S$,  
with the final state, described by two wave functions. 
One of them is the
eigenfunction $\Psi_t^{s_t}$, with eigenvalue
$E = E_{min}+E_R^*$, of the state $s_t$ of the intrinsic
Hamiltonian pertaining to the system of two {\sl interacting}
nucleons with relative momentum $\vec t$, 
which can be either
a bound 
or a scattering state. The other one
is
the plane wave representing 
the nucleon $N$ in IA.

The overlap Eq.(\ref{trueover}) is therefore 
the crucial nuclear ingredient to evaluate the
light cone correlator, Eq.(1), for $^3$He.}
Let us show how it is obtained through
the quantity at our disposal, which is actually the following:
 
\begin{eqnarray}
\tilde{{ I}}^{X, j_{12}, \eta }_{L_{\rho} {\tau},\Sigma}  (y_1 ,E) & =&
\int d\Omega_{\hat y_1} ~ 
\sum_{M_{\rho},M_X} \langle X M_X, L_{\rho} M_{\rho}|{1 \over 2}J_z \rangle  ~
{Y}^*_{L_{\rho} M_{\rho}} (\Omega_{\hat y_1} )\left [{\cal T}_{1 \over 2}^{\tau}\right ]^{\dagger} ~
\nonumber \\
&\times &\sum_{m_{12}\sigma}  
\langle j_{12}m_{12}, {1 \over 2} \sigma|XM_X \rangle   
\langle {\bf y}_1, {1 \over 2}\sigma;\psi^{j_{12},m_{12},\Sigma}_
{E,\eta}  |\Psi_{^3He}^{{1 \over 2}J_z}\rangle \ .
\label{overcon}
\end{eqnarray}
Here above,
$\rho$ denotes quantum numbers of the 
struck nucleon,  treated as a
free particle in IA, with spin projection $\sigma$, 
isospin projection $\tau $, orbital angular momentum and its third
component $L_{\rho}, ~ M_{\rho}$, respectively. 
$ \psi^{j_{12},m_{12}\Sigma}_{E,\eta} $ represents the two-body recoiling
system with total angular momentum and its third component $j_{12}, ~ m_{12}$, 
spin $\Sigma$ and
excitation energy $E$. 
$\eta$ is an additional quantum number necessary to describe the 
OAM mixing in the interacting
two-body 
state in the continuum.
$ j_{12} $ is coupled to the spin $\frac{1}{2}$ of the interacting nucleon 
to give the intermediate
angular momentum $X$. 
The description of the three-body system is based on the Av18 
calculation of the wave function 
of Ref. \cite{tre},  given in terms of the
following Jacobi coordinates ${\bf y}_1$ and ${\bf x}_1$:

\begin{eqnarray}
{\bf y }_1 &=&-{2 \over \sqrt{ 3}}~\left [\bf{r}_1- {\bf{r}_2+\bf{r}_3 \over 2}
 \right ]=-{2 \over \sqrt{ 3}}~{\bm \rho}
 \nonumber \\
 {\bf x }_1 &=&\left [\bf{r}_2- \bf{r}_3
 \right ]={\bf r}
\end{eqnarray}

\noindent
where ${\bf r}_i$ (with $i=1,2,3$) are the CM variables of the nucleons
in the trinucleon, constrained by 
${\bf r}_1+{\bf r}_2+{\bf r}_3=0$.
The trinucleon bound state, 
$ \Psi_{^3He}^{\frac{1}{2}J_z}({\bf x}_1,{\bf y}_1) $, can be 
written in terms of the variables  
${\bf y }_1$ and ${\bf x }_1$ using the
following basis for the isospin-spin-angular part:

\begin{eqnarray}
{\cal Y}^{J_{12} \Sigma \ell_{12}}_{L_{\rho}XJJ_z}(\hat{x}_1,\hat{y}_1) &=& \sum_{M_{\rho} M_X}
\sum_{m_{12} \sigma}
\langle j_{12} M_{12}, {1 \over 2} \sigma|X M_X \rangle  
\nonumber  \langle X M_X,  L_{\rho} M_{\rho} 
|\frac{1}{2}J_z \rangle 
\nonumber \\
&\times &\sum_{m_{12}}\sum_{\mu}~\chi_{\Sigma}^{\mu}~{Y}_{ \ell_{12} m_{12}}
( \Omega_{{ \hat  x}_1})~\langle \ell_{12} m_{12}, {\Sigma} \mu|j_{12} M_{12} \rangle ~ 
\nonumber \\
&\times &\sum_{ \tau \tau_{12}} 
\langle {1 \over 2} \tau, T_{12} \tau_{12} |\frac{1}{2}\frac{1}{2} \rangle ~
{\cal T}_{1 \over 2}^{\tau}~{\cal T}_{T_{12}}^{\tau_{12}}{Y}_{L_{\rho}M_{\rho}}
( \Omega_{{ \hat  y}_1}) \chi_{1 \over 2}^{\sigma}
\label{basis}
\end{eqnarray}
\noindent
through the radial overlaps 
\begin{eqnarray}
\tilde{{\cal I}} ^{j_{12}\Sigma \ell_{12}}_{L_{\rho}X}
(
|{\bf x}_1|
,
|{\bf y}_1|
) = 
\int d\Omega_{{ \hat  y}_1}
\int d \Omega_{{ \hat  x}_1}
\left [{\cal Y}^{j_{12}\Sigma\ell_{f}}_{L_{\rho}XJJ_z}(\hat{x}_1,\hat{y}_1) \right ]^{\dagger}
\Psi_{^3He}^{\frac{1}{2}J_z}({\bf x}_1,{\bf y}_1)
\label{over}
\end{eqnarray}
\noindent
with the constraints
\begin{eqnarray}
&&( -1 )^{\ell_{f} + L_{\rho}} = 1~,
\nonumber \\
&&( -1 )^{\ell_{f} + \Sigma + T_{12}} = -1  \; \;  ,
\label{const}
\end{eqnarray}
\noindent
following from the parity of the nucleus and 
the Pauli principle, respectively.

\noindent

The quantity at our disposal, Eq. (\ref{overcon}), can be defined
in terms of the overlaps Eq. (\ref{const}), as follows:

\begin{eqnarray}
\tilde{{ I}}^{X, j_{12}, \eta }_{L_{\rho} {\tau},\Sigma}  (y_1 ,E) & =&
 \langle {1 \over 2} \tau, T_{12} \tau_{12} |\frac{1}{2} \frac{1}{2} \rangle ~
\sum_{\ell_{12}}  ~
\int ~ {\tilde{{\cal I}}} ^{j_{12}\Sigma \ell_{12}}_{L_{\rho}X}(|{\bf y}_1|,|{\bf x}_1|) ~ 
u^{j_{12},E,\eta}_{ \ell_{12}}(x_1) ~ x_1^2 ~ dx_1~,
\label{tilde} 
\end{eqnarray}
\noindent
where
$ u^{j_{12},E,\eta}_{ \ell_{12}}(x_1) $ is the two-nucleon radial function.

Using this formalism
to calculate the spectral function, Eq. (\ref{spectral1}), one has to obtain
the product of two intrinsic overlaps, integrated over the angles of 
$\vec t$,
the 
relative momentum of the recoiling pair. This quantity is expressed 
through the overlaps
Eq. (\ref{tilde}) which, as previously said, are the quantities at disposal,
as follows:
\begin{eqnarray}
\label{overlap}
&&\int d\Omega_t\langle\vec{P}'S'|\vec{t}s_t,\vec{p}\,'s'\rangle    
\langle \vec p \, s, \vec t \, s_t| \vec P \, S\rangle  =
\nonumber \\
&=& \int d\Omega_t \int d\rho' e^{-i\vec{p}\,'\vec{\rho'}}
\int d\rho e^{i\vec{p}\vec{\rho}}\langle  \rho ,\frac{1}{2}s;\psi^{j_{12},m_{12}\Sigma}_{E,\eta}|\Psi^{\frac{1}{2}J_z,\frac{1}{2}T_z}_{^3He}\rangle
\nonumber \\
&\times &\langle \Psi_{^3He}^{\frac{1}{2}J'_z,\frac{1}{2}T_z}|\psi_{E,\eta}^{j'_{12},m'_{12}},\frac{1}{2}s',\rho'\rangle \left[\tau^{\tau}_{\frac{1}{2}}\right]^{\dagger}\left[\tau^{\tau}_{\frac{1}{2}} \right] =
\nonumber \\
&=& \int d\rho' e^{-i\vec{p}\,'\vec{\rho'}}\int d\rho e^{i\vec{p}\vec{\rho}}\sum_{X,M_X} \sum_{X',M_{X'}}\sum_{m_{12},s}\sum_{s'}
\mathit{Y}^{\ast}_{L,M}(\Omega_{\rho}) \mathit{Y}_{L',M'}(\Omega_{\rho'})
\nonumber \\
&\times & \langle X M_X,L,M | \frac{1}{2} J_z \rangle \langle j_{12}m_{12}, \frac{1}{2}s|XM_X\rangle\langle J'_z \frac{1}{2}|L'M',X'M_{X'}\rangle
\nonumber \\
&\times &\langle X'M_{X'}|\frac{1}{2}s',j_{12}'m_{12}\rangle \tilde{I}^{\Sigma ,X,j_{12},T_{12},\zeta}_{L,\tau}(\rho , E)\tilde{I}^{\Sigma ,X',j_{12},T_{12},\zeta}_{L',\tau}(\rho' , E)
\end{eqnarray}
\noindent
Using these relations, together with Clebsch-Gordan coefficients 
and spherical
harmonics properties, the final expressions for the relevant components
of the light cone correlator are found:

\begin{eqnarray}
\label{finalf++}
F_{++}^{3,q,0}(x,\xi,\Delta^2)&=& \dfrac{M}{(2\pi)^5}\sum_{N}\int_0^{\infty}dE\sqrt{ME} \int_0^{\infty} dp \ p^2 \sum_{L_{\rho},j_{12},X,\Sigma}
\nonumber \\
&\times &\int d\rho \ \rho^2 j_{L_{\rho}}(p\rho)\tilde{I}_{L_{\rho},\tau}^{\Sigma,X,j_{12},T_{12}}(\rho,E) 
\nonumber \\
&\times &\int_{-1}^{1}d\mathit{cos\theta}_p\int_{0}^{2\pi}d\phi_p P_{L_{\rho}}(\mathit{cos\theta}_{(\vec{p}+\vec{\Delta}\widehat{)}\vec{p}})
\nonumber \\
&\times &\int d\rho' \rho'^2 j_{L_{\rho}}(|\vec{p}+\vec{\Delta}|\rho')\tilde{I}_{L_{\rho},\tau}^{\Sigma,X,j_{12},T_{12}}(\rho',E)F_{++}^{N,q,0}(x',\xi',\Delta^2) \ .
\end{eqnarray}

\begin{eqnarray}
 F_{+-}^{3,q,1}(x,\xi,\Delta^2)  &=& \dfrac{M}{(2\pi)^4} \sum_{N}\sum_{\substack{\gamma'}} \int dE \dfrac{\sqrt{ME}}{2}\int d\vec{p} 
\int d\rho \rho^2 \int d\rho'\rho'^2 
\nonumber \\
&\times &j_{L_{\rho'}}(|p+\Delta |\rho ')j_{L_{\rho}} (p\rho) 
\tilde{I}_{\substack{\beta'}}(\rho',t)\tilde{I}_{\substack{\beta}}(\rho,t)\mathit{Y}_{L_{\rho}M_{\rho}}(\Omega_{p})(-1)^{L_{\rho'}+\frac{L_{\rho'}+L_{\rho}}{2}}
\nonumber \\
&\times &(F_{\substack{\varrho}}\mathit{Y}^{\ast}_{L_{\rho'},M_{\rho}-2}(\Omega_{p+\Delta}) -D_{\substack{\varrho}} \mathit{Y}^{\ast}_{L_{\rho'},M_{\rho}}(\Omega_{p+\Delta}))F_{+-}^{N,q,1}(x',\xi',\Delta^2)   
\end{eqnarray}
\noindent
where $ \gamma' = L_{\rho},M_{\rho},X,j_{12},\Sigma,\lambda,L_{\rho'},X' $.
$F_{\substack{\varrho}}$ and $D_{\substack{\varrho}}$ are terms containing  
the product
of four Clebsch-Gordan coefficients, depending on the set of quantum numbers 
\noindent
$\varrho = L_{\rho'} L_{\rho},M_{\rho} ,X,X',J_{12}$.

Using the latter two equations in Eqs. (\ref{relazioni}), the nuclear GPDs,
Eqs. (\ref{H})
and (\ref{HpE}), can be written 
through the overlaps at our disposal and,
eventually, numerically evaluated. 

\twocolumngrid

\end{document}